\documentstyle[aps]{revtex}

\def\be{\nopagebreak[3]\begin{equation}}
\def\ee{\end{equation}}
\def\ba{\nopagebreak[3]\begin{eqnarray}}
\def\ea{\end{eqnarray}}

\newcommand{\teta}{\rlap{\lower2ex\hbox{$\,\tilde{}$}}\eta{}}

\begin{document}
\title{Comments on Challenges for Quantum Gravity}
\author{ Alejandro Perez${}^1$ and Daniel Sudarsky${}^{1, 2}$ }
\address{1. Center for Gravitational Physics and Geometry\\
Pennsylvania State University\\
University Park, PA 16802, USA  \\
}\address{2. Instituto de Ciencias Nucleares\\
Universidad Nacional Aut\'onoma de M\'exico\\
A. Postal 70-543, M\'exico D.F. 04510, M\'exico\\
}

\maketitle

\begin{abstract}
We examine radiative corrections arising from Lorentz violating dimension five operators presumably associated with
Planck scale physics as recently considered by Myers and Pospelov. We find that observational 
data result in bounds on the dimensionless parameters of the order $10^{-15}$. These represent the most stringent bounds 
on Lorentz violation to date. 
\end{abstract}





There has been recently a great deal of interest in possible
modifications of the dispersion relations for ordinary particles
that might  be the result of
 quantum gravitational effects \cite{Quant Fluct}. These effects are thought
to arise from a breakdown of Lorentz Invariance  and would imply,
in contrast with one of the most cherished and useful principles
of physics, the existence of
  a preferential frame, a new version of the XIX${}^{th}$ century Ether.
In fact a large  collection of   ever tighter bounds have been
obtained by considering astrophysical observations and Laboratory
experiments. In a recent letter Myers and Pospelov \cite{myers} have considered
such  phenomena in the language of effective field theory
 describing such effects in terms of  Lorentz Violating
dimension five operators in the  Lagrangian for free fields.
 In fact, some of these terms are obtained in heuristic
models suggested by Loop Quantum Gravity.  We want to touch upon
three aspects of the discussion: First that the mechanism
suggested by the authors to avoid the appearance of new
unsuppressed terms trough radiative corrections does not work
beyond the linear order in the dimensionless parameters $\xi,\eta_1,$ and $\eta_2$ below. Second that the idea that the
natural cutoff for such terms might be low, say the SUSY scale ($\approx TeV$), conflicts with the basic
 underlying rational for
assuming the existence of the effects. And third, to point out
that although these results could be seen as essentially ruling
out the heuristic models, they are not, as of now, saying anything
about the theory of Loop Quantum Gravity besides limiting the
degree to which its relevant semiclassical states 
might break Lorentz Invariance over a macroscopic regime.

We will focus  attention on the treatment of \cite{myers}, particularly, 
the terms corresponding to the
gauge bosons fields  $A_{\mu}$ and the fermion  fields $\Psi$,
\be 
{\cal L}_{\gamma}=\frac{\xi}{M_p} C^{a b c}\epsilon^{d}_{\ \, bef} \  F_{ad} \partial_{b} F^{ef}
\ee
 and 
\be
{\cal L}_{f}=\frac{1}{M_p}\bar \Psi C^{abc}(\eta_1 \gamma_{a}+\eta_2 \gamma_a \gamma_5) \partial_b \partial_c \Psi
\ee
respectively, and where $C^{abc}=W^{a}W^{b}W^{c}$ with $W^{a}$ the 4-velocity of the
preferential frame. These lead to corrected free propagators which
can be written as:
\be \Delta_{\gamma}^{ab}(k)= -i \left[g^{ab} k^2 + i \frac{\xi}{M_p} \ \epsilon^{abcd}C_{def}k_{c}k^{e}k^{f} \right]^{-1}\ee
for the photon, and
\be \Delta_{f}(p)=i\left[ p_{a}\gamma^{a}-m -\frac{1}{M_p}
C^{abc} (\eta_1 \gamma_{a}+\eta_2 \gamma_a \gamma_5) p_b p_c \right]^{-1}\ee
for the fermions. The issue is now, what is the effect of using these
corrected propagators in the self energy of the fermion? Myers et al. \cite{myers}
noted that these would lead to a generation of a large Lorentz
 violating effects represented by dimension 2 or 3 operators. 
They then suggest
modifying the scheme by replacing the tensor $C^{abc}$
 by the tensor $\tilde C^{abc}=W^a W^b W^c -(1/6)(W^a \eta^{bc}+ W^b 
\eta^{ac}+W^c\eta^{ab})$
 which has the property that  it vanishes on  contraction with
the flat Minkowski tensor $\eta_{ab}$ in any pair of indices.
This feature then ensures the integrals such as $\int dk^4 \frac{k_{\mu}k_{\nu}}{k^4}\propto \eta_{\mu\nu} \Lambda^2$
(where $\Lambda$ is the cut-off of the effective theory),
appearing in the calculation of the self energies,
will not result in large  Lorentz violating terms. Our first point is
that when one goes beyond the lowest order in $\xi$, $\eta_1$ and $\eta_2$ (in particular beyond second order) 
one finds integrals such as 
$\int dk^4 \frac{k_{\mu}k_{\nu} k_{\rho} k_{\sigma}k_{\tau} k_{\alpha} k_{\beta} k_{\gamma}}{k^8} \propto \Lambda^4 
(\eta_{\mu\nu}\eta_{\rho \sigma}\eta_{\tau \alpha}\eta_{\beta \gamma}+ \ {\rm perm.})$, 
which when contracted with $C_{\mu \nu \rho}C_{\sigma \tau \alpha}C_{\beta \gamma \delta}$
results in a non-vanishing term proportional to $W_{\delta}$.

These then generate the dangerous low dimension operators that one was
trying to avoid. In particular the  one loop self energy of a
charged fermion  generates the following
 Lorentz Violating effective term:
\be
e^2 \frac{\Lambda^4}{M_p^3}P^{(3)}(\xi, \eta_1, \eta_2) \ \bar \Psi W^{a}\gamma_{a}\gamma_{5}\Psi \label{tanto}
\ee 
where $e$ is the electromagnetic coupling and $P^{(3)}$ is a
polynomial   of degree $3$ with coefficients of order $1$.
This example shows that the well known expectation from
effective field theories, namely that all operators allowed by the
remaining symmetries would be generated, with coefficients
 of the appropriate order in the cut-off scale, unless the theory is
 renormalizable, can not be overcome by  a simple recipe for the detailed
form of the terms.

Similarly, considering the  vacuum polarization
one generates a Chern Simon type term in the photon propagator
\be 
e^2 \frac{\Lambda^4}{M_p^3}P^{\prime (3)}(\eta_1, \eta_2) \ \epsilon^{abcd} W_{a}A_{b}F_{cd}\label{cs}
\ee
where $P^{\prime (3)}(\eta_1,\eta_2)$ is also a polynomial of degree $3$.

Next we consider the idea mentioned in \cite{myers} to set the cutoff scale for the
effective theory at a very low value such as the scale normally
attributed to the supersymmetry breaking. 
This would mean that the phenomena of interest, having its origins at the
Planck scale, somehow is  hidden by some mechanism,  that 
effectively protects Lorentz invariance  from large
 violations (low dimensional operators) throughout the so called ``scale
dessert" from $10^3 Gev $ to $10^{19}Gev$. Let's recall that the underlying hypothesis
behind these
 considerations is that Planck scale physics is directly connected 
with
the low energy physics scale  becoming accessible in some
astrophysical phenomena. Thus, 
 while something like this is in principle conceivable, in its consideration,
a careful reexamination of the  whole scenario would seem to be
required. In particular
simple supersymmetry  does not seem to be sufficient to eliminate the renormalizable terms described here. 

The above discussion concerning consistency with the idea that the Planck 
scale is becoming accessible in these experiments indicates that one should set the cut-off
for the effective theory around the Planck scale ($\Lambda \approx M_p$). 
The fermion corrections in equation (\ref{tanto}) are tightly bounded experimentally.
Direct comparison of (\ref{tanto}) with equation (9) in \cite{ustedes} results in the remarkable bound
$P^{(3)}(\xi, \eta_1, \eta_2) < 10^{-45}$. This
lead us to conclude that the parameters $\xi, \eta_1$, and, $\eta_2$ 
must be at most of order $10^{-15}$. Similarly, a comparison of the effects of equation (\ref{cs})
with those studied in \cite{kozameh} leads to $P^{\prime (3)}(\eta_1,\eta_2) < 10^{-57}$ indicating that 
the parameters $\eta_1$, and $ \eta_2$ must be at most of order $10^{-19}$. We should point out that while the
 last bound could in fact be dramatically lowered by introduction of the hypothesis of supersymmetry,
 the former (and weaker)  one seems to be very robust. 
These bounds dwarf any bound extracted from the analysis of astrophysical
phenomena or that could be set by any gamma ray burst 
experiment in the foreseeable future.

Finally, we must emphasize that at this point one is not testing the theory
of Loop Quantum Gravity, as nothing in this framework necesitates the kind of breakdown
of Lorentz Invariance that is associated with the existence of a preferential reference frame. 
All one has at this point are heuristic proposals for states of the theory \cite{LoopQG}, 
that would result in the types of 
effects discussed in \cite{Quant Fluct}.
Moreover, given any of such state,  there is nothing in principle
preventing the construction of new states by applying a Lorentz boost 
(using an appropriate Loop quantum gravity operator) to the original state.
In this way one could conceive a suitable superposition of states which 
will not be associated with any preferential frame. Such type of scenario 
would be immune to the constraints being set by the current explorations
of Quantum Gravity Phenomenology.

\section*{Acknowledgments}
We would like to thank Murat Gunaydin for helpful discussions.
D.S. would like to acknowledge partial support from DGAPA--UNAM
Project No IN 112401, a CONACYT Sabbatical Fellowship and 
the hospitality of the CGPG. This work was supported in part by 
NSF grant PHY00-90091 and the Eberly research funds of Penn State.


\begin{references}

\bibitem{Quant Fluct}  G. Amelino-Camelia, J. Ellis, N.E. Mavromatos, D.V.
Nanopoulos and S. Sarkar, {\ Nature} (London) {\bf 393}, 763
(1998);, {Nature} (London) {\bf 400}, 849 (1999); D.V. Ahluwalia,
Nature (London) {\bf 398}, 199 (1999); G. Amelino-Camelia, Lect.\
Notes Phys.\ {\bf 541}, 1 (2000).

\bibitem{myers} R.C. Myers and M. Pospelov, hep-ph/0301124, (to appear in Phys. Rev. Lett.)

\bibitem{ustedes} D. Sudarsky, L. Urrutia, and H. Vucetich, {\ Phys. Rev. Lett.}, {\bf 89}, 231301 (2002).

\bibitem{kozameh} R.J. Gleiser, and C.N. Kozameh, {\ Phys. Rev.}, {\bf D64}, 083007 (2001).

\bibitem{LoopQG}  R. Gambini and J. Pullin, {\ Phys. Rev.} {\bf D59}, 
124021 (1999), J. Alfaro, H. Morales-T\'ecotl and L. Urrutia, {\ Phys. Rev.}
{\bf D65}, 103509 (2002); J. Alfaro, H. Morales-T\'ecotl and L.
Urrutia, {\ Phys. Rev. Lett.} {\bf 84}, 2318 (2000).

\end{references}
\end{document}